\documentclass[aps,prx,twocolumn,floatfix,longbibliography,superscriptaddress,10pt,showkeys]{revtex4-1}

\usepackage[english]{babel}

\usepackage[utf8]{inputenc}
\usepackage[T1]{fontenc}

\usepackage{amsthm}
\usepackage{amssymb}
\usepackage{mathtools}
\usepackage{units}
\usepackage{braket}

\usepackage{xcolor}
\usepackage[pdftex]{graphicx}

\usepackage{hyperref}
\hypersetup{colorlinks=true, citecolor=blue, urlcolor=blue, linkcolor=blue}

\newcommand{\expect}[1]{ \braket{#1}} % Expectation value
\newcommand{\operator}[1]{\hat{#1}}

\newcommand{\cone}{\mathrm{i}}
\newcommand{\half}[1]{\ensuremath{\mathtt{#1}}}
\renewcommand{\vec}[1]{\boldsymbol{#1}}
\newcommand{\angstrom}{\textup{\r{A}}}

\begin{document}
\title{Nonlinear spin and orbital Rashba-Edelstein effects induced by a femtosecond laser pulse: Simulations for Au(001)}

\author{Oliver Busch}
\author{Franziska Ziolkowski}
\author{B{\"o}rge G{\"o}bel}\email[Correspondence email address: ]{boerge.goebel@physik.uni-halle.de}
\author{Ingrid Mertig}
\author{J{\"u}rgen Henk}
\affiliation{Institut f{\"u}r Physik, Martin Luther University Halle-Wittenberg, 06099 Halle, Germany}

\date{\today}

\begin{abstract}
Rashba-type spin-orbit coupling gives rise to distinctive surface and interface phenomena, such as spin-momentum locking and spin splitting. In nonequilibrium settings, one of the key manifestations is the (Rashba-)Edelstein effect, where an electric current generates a net spin or orbital polarization perpendicular to the current direction. While the steady-state behavior of these effects is well studied, their dynamics on ultrafast timescales remain largely unexplored. In this work, we present a theoretical investigation of the ultrafast spin and orbital Edelstein effects on an Au(001) surface, triggered by excitation with a femtosecond laser pulse. These effects are intrinsic and inherently nonlinear. Using a real-space tight-binding model combined with time evolution governed by the von Neumann equation, we simulate the electron dynamics in response to the pulse. Our results reveal pronounced differences between the spin and orbital responses, offering detailed insights into their distinct temporal profiles and magnitudes. We further explore the associated charge, spin, and orbital currents, including the emergence of laser-induced spin and orbital Hall effects. Finally, we quantify the angular momentum transfer mediated by the light-matter interaction. These findings shed light on the intricate ultrafast dynamics driven by spin-orbit coupling and offer guidance for the design of next-generation spintronic and orbitronic devices.

\end{abstract}

%\keywords{Condensed matter physics, ultrafast electron dynamics, Nonlinear Rashba-Edelstein effect, Hall effect, tight-binding model, numerical simulations}

\maketitle

\section{Introduction}\label{sec:introduction} 
Rashba-type spin-orbit coupling (RSOC) manifests itself in a number of effects at surfaces or interfaces~\cite{bychkov1984, Bihlmayer2022}. A prominent example is the spin-momentum locking of surface states, for example, in the L-gap surface state in Au(111)~\cite{Henk2004}, and the accompanying spin splitting, for example, in the surface alloy Bi/Ag(111)~\cite{Ast2007}. Concerning nonequilibrium situations, the (Rashba-)Edelstein effect is a clear signature of RSOC~\cite{Aronov1989, edelstein1990, inoue2003, gambardella2011, johansson2024}: an electric current flowing parallel to a surface produces a homogeneous spin polarization in the surface layers that is in-plane and perpendicular to the current's direction. This spin Edelstein effect (SEE) is complemented by the orbital Edelstein effect (OEE), in which the current results in an orbital polarization~\cite{zhong2016, go2017, yoda2018, salemi2019, johansson2021, johansson2024, goebel2025}.
 
Both the spin and the orbital Edelstein effect are well understood for the steady state in which the applied electric field is time-independent. For ultrafast electron dynamics, results obtained for the steady state cannot be easily transferred into the time domain. An example is the ultrafast orbital Hall effect generated by a femtosecond laser pulse, which exhibits similarities but also striking differences in comparison to its steady-state counterpart~\cite{busch2024}. 

Concerning ultrafast Edelstein effects, the following questions arise. 
Are there significant differences between the SEE and the OEE? Recall that spin is an inherent property of an electron, whereas its orbital moment is related to its motion. Are the ultrafast Edelstein effects significant mainly `during the laser pulse' or is there a sizable signal `after the pulse'? Are the Edelstein effects accompanied by currents of spin and orbital angular momentum, either in longitudinal or transversal directions (i.\,e.\ Hall currents)? How large are the magnitudes of the laser-induced spin angular momentum (SAM) and orbital angular momentum (OAM)?

\begin{figure}
    \centering
    \includegraphics[width=1.0\columnwidth]{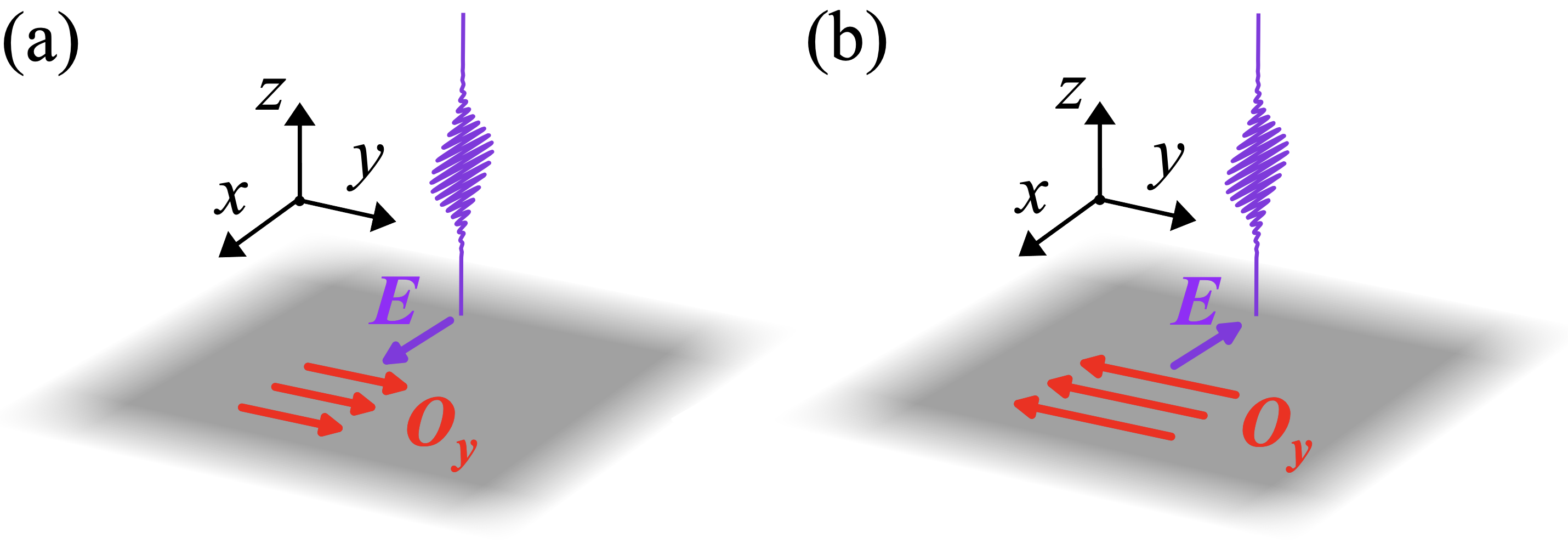}
    \caption{Sketch of the ultrafast spin and orbital Edelstein effect in a two-dimensional system. (a) A linearly polarized laser pulse with electric field $E$ along the $+x$-direction (violet) impinges perpendicular to the sample (gray; located in the $xy$ plane). $O_y$ (red) is the laser-induced in-plane angular momentum $O$ (either spin or orbital angular momentum) along the $+y$-direction. (b) Same as (a) but half a period later: $E$ and $O_y$ point along the $-x$- and the $-y$-direction, respectively.}
    \label{fig:sketch}
\end{figure}

In this paper, we give answers to the above questions. We report on a theoretical study of the SEE and the OEE generated by a femtosecond laser pulse. As a sample, we choose an Au(001) surface, which is subject to RSOC; the latter may be induced by an insulating substrate. The sample is illuminated by a femtosecond laser pulse (Figure~\ref{fig:sketch}), which excites the electron system via dipole transitions. The electron dynamics is described by a one-electron density matrix, the time evolution of which is given by the von Neumann equation. The Edelstein effects are then analyzed in terms of time-dependent spin and orbital polarizations as well as in terms of their respective currents flowing within the Au sample.

Ultrafast phenomena can be delineated on the level of time-dependent density functional theory (TDDFT), a very accurate description~\cite{Ullrich2011, Krieger2015, Dewhurst2018}. Being computationally demanding, TDDFT simulations are usually restricted to small sample sizes and short time spans. By contrast, model calculations, which require less computational effort, may not capture important details of the electronic structure, in particular, if highly excited electronic states are involved in the dynamics. Our computational approach \textsc{evolve}~\cite{Toepler2021, Ziolkowski23, busch2023a, busch2023b, busch2024} bridges the gap between TDDFT and model calculations: based on a tight-binding method in real space, it allows treating comparably large samples and long time spans, and it represents the electronic structure reasonably well. \textsc{evolve} has proven its suitability in a number of investigations. For example, a backflow mechanism in Co/Cu heterostructures is faithfully reproduced~\cite{Toepler2021}. The tight-binding description imposes two restrictions: the discrete representation in terms of sites and atomic orbitals imposes, first, that spatial resolution is limited to atomic sites and, second, the hopping matrix elements in the Hamilton operator introduce a lower limit of the temporal resolution (in the order of a few tens of attoseconds).

This paper is organized as follows. Theoretical aspects are sketched in Section~\ref{sec:theoretical_aspects}. Section~\ref{sec:results} collects the discussion of the results, and concluding remarks are given in Section~\ref{sec:conclusion}. In the Appendix, we present the symmetry analysis (\ref{sec:symmetry}), derive the RSOC in a real-space tight-binding approach (\ref{sec:Rashba-SOC-TB}), and present results for a sample without RSOC (\ref{sec:noRSOC}). A frequency analysis provides a means to distinguish SAM and OAM currents with regard to their origin (\ref{sec:frequencyAnalysis}).

\section{Theoretical aspects}
\label{sec:theoretical_aspects} 
We briefly introduce the main ideas of our approach to ultrafast electron dynamics. The computations have been performed with the computer code \textsc{evolve}, which is being developed in our group (for details, see Refs.~\onlinecite{Toepler2021, Ziolkowski23, busch2023a, busch2023b, busch2024}).

The samples are Au(001) monolayers. The free-standing films form a square lattice, with Cartesian axes chosen as $x \equiv [110]$, $y \equiv [\bar{1}10]$, and $z \equiv [001]$. The samples are approximated by real-space clusters with dimension $11 \times 11$ atomic sites, with periodic boundary conditions applied in $x$- and in $y$-directions (i.\,e.\ closed circuit geometry). Au lends itself to a study of the ultrafast SEE and OEE since it exhibits a sizable spin Hall conductivity~\cite{Seki2008, Isasa2025, Zou2016} and a moderate Rashba spin-splitting~\cite{Lashell1996, Nicolay2001, Hoesch2004, Cercellier2006}.

The electronic structure of the samples is described by a semi-empirical real-space tight-binding (TB) approach~\cite{Slater1954}. The Hamiltonian $\operator{H}_{0}$ is of Slater-Koster-type with parameters taken from Ref.~\onlinecite{Papaconstantopoulos2015}.
\textcolor{black}{The difference of the on-site energies of the $p$- and $d$-orbitals is about $\unit[14]{eV}$, and the hopping integrals have a magnitude of maximal $\unit[2.5]{eV}$. $\operator{H}_{0}$} includes both the atomic spin-orbit coupling \textcolor{black}{(SOC)}~\cite{Konschuh2010} and the RSOC for gold~\cite{Barreteau2016}. The latter mimics the effect of a substrate which induces a potential gradient perpendicular to the film; for details, see Appendix~\ref{sec:Rashba-SOC-TB}. 
\textcolor{black}{Values for the atomic SOC strength of Au have been taken from Ref.~\onlinecite{Barreteau2016} and read $\xi_p = \unit[1.5]{eV}$ and $\xi_d = \unit[0.65]{eV}$ for $p$- and $d$-orbitals, respectively. We have estimated the RSOC parameter $\lambda^R_{z\alpha}$ to about $\unit[0.1]{eV}$ with $\gamma^\mathrm{SO} = \unit[4.2\cdot10^{-9}]{eV\cdot cm}$ from Ref.~\onlinecite{Barreteau2016} and with the Au lattice constant of $\unit[4.08]{\angstrom}$.}

The electronic system described by $\operator{H}_{0}$ is excited by a femtosecond laser pulse~\cite{Savasta1995}, whose electric field oscillates within the sample along the $x$-axis. This field has a sinusoidal carrier wave with an energy of $\hbar \omega = \unit[3.0]{eV}$ (equivalent to a period $T$ of about $\unit[1.38]{fs}$) and a Lorentzian envelope of $\unit[10]{fs}$ width and center at $t = \unit[0]{fs}$. The photon energy was chosen to allow excitation of the occupied $d$-states (centered about $\unit[3]{eV}$ below the chemical potential) into the unoccupied $p$-orbitals. The laser's amplitude is chosen to obtain a fluence of about~$\unit[3.2]{mJ\,cm^{-2}}$ and $0.15$ excited electrons per site (at the laser's maximum; about $0.05$ thereafter). The geometry of the entire setup (sample and laser) dictates that only certain components of the SAM and the OAM are produced by the incident radiation (see Refs.~\onlinecite{Henk1996,busch2023a,busch2023b, busch2024}, as well as the symmetry analysis in Appendix~\ref{sec:symmetry}).

The ultrafast electron dynamics is described by the von Neumann equation
\begin{align}
   -\cone \hbar \frac{\mathrm{d} \operator{\rho}(t)}{\mathrm{d} t} & =  [ \operator{\rho}(t),\operator{H}(t)]
    \label{eq:EOM}
\end{align}
for the one-particle density matrix $\operator{\rho}(t)$. The latter is either expressed in a site-orbital-spin basis or in the eigenstate basis of $\operator{H}_{0}$. The time-dependent Hamiltonian $\operator{H}(t)$ supplements $\operator{H}_{0}$ by the rapidly oscillating laser radiation. The latter is taken into account by a unitary transformation~\cite{Savasta1995} which includes dipole transitions in first order in the electric field.

Since we are interested in spatial-temporal signatures of SAM, OAM, and their currents, we express the density matrix $\operator{\rho}(t)$ in the site-orbital-spin basis. This allows us to compute expectation values $\expect{O}(t) = \operatorname{tr}[\operator{\rho}(t) \,\operator{O}]$, in which the trace can be restricted to achieve site, orbital, and spin resolution. We are particularly interested in the occupation probabilities 
\begin{align}
    \expect{p_{k}}(t) & = \operatorname{tr}_{k}[\operator{\rho}(t)],
    \label{eq:occupation}
\end{align}
as well as in the spin and orbital polarization
\begin{subequations}
\begin{align}
    \expect{\vec{s}_{k}}(t) & = \operatorname{tr}_{k}[\operator{\rho}(t) \,\operator{\vec{\sigma}}],
    \label{eq:SAM}
    \\
    \expect{\vec{l}_{k}}(t) & = \operatorname{tr}_{k}[\operator{\rho}(t) \,\operator{\vec{L}}]
    \label{eq:OAM}
\end{align}
\end{subequations}
at site $k$ ($\operator{\vec{\sigma}}$ vector of SAM operators, $\operator{\vec{L}} = (\operator{L}_{x}, \operator{L}_{y}, \operator{L}_{z})$ vector of OAM operators). Orbital momenta are calculated in the atomic center approximation (ACA)~\cite{go2017, pezo2022, busch2024} and are brought about by hybridization of orbitals~\cite{go2018, go2021}. 

The electric field of the laser does not only change the above quantities locally but also results in respective currents~\cite{mahan2000,busch2023a, busch2024}. The probability current from site $l$ to site $k$ is calculated from (in atomic units)
\begin{align}
	\expect{j_{kl}}(t) & \equiv 
    \frac{\mathrm{i}}{2} \expect{\rho_{lk}(t) \, h_{kl}(t)}
    - \expect{l \leftrightarrow k}.
    \label{eq:current}
\end{align}
$\rho_{lk}$ and $h_{kl}$ are off-diagonal blocks of the density matrix and of the Hamiltonian matrix in the site representation, respectively. Transport of the $\mu$-th momentum component from site $k$ to site $l$ is quantified by~\cite{busch2023b, busch2024}
\begin{align}
	\expect{j_{kl}^{O_\mu}}(t) \equiv \frac{1}{2} \left[ \expect{O_{k}^{\mu} j_{kl}}(t) + \expect{j_{kl} O_{l}^{\mu}}(t) \right], \mu = x,y, z.
    \label{eq:SAM-OAM-current}
\end{align}
For the latter, $O_{k}^{\mu}$ and $O_{l}^{\mu}$ are either a SAM or an OAM operator taken with respect to the sites' positions.

A typical simulation performed with \textsc{evolve} proceeds as follows. In the first step, the Hamiltonian $\operator{H}_{0}$ is set up and diagonalized, which allows us to determine the chemical potential of the ground state and the initial occupation probabilities. Subsequently, the von Neumann equation~\eqref{eq:EOM} is solved, starting at least $\unit[50]{fs}$ before the laser pulse's maximum (depending on the pulse width). At each time $t$, the site-resolved expectation values of the observables, equations~\eqref{eq:occupation}--\eqref{eq:SAM-OAM-current}, are computed. 

\section{Results and discussion}
\label{sec:results}
A photo excitation of an electron system produces, in general, SAM and OAM, an effect well-known from photoelectron emission spectroscopy~\cite{Tamura87, Tamura91a, Tamura91b, Henk1994, Henk1996, uenzelmann2020, Kobayashi2020}. We performed a symmetry analysis in order to identify the RSOC-induced SAM and OAM components.

The symmetry analysis, detailed in Appendix~\ref{sec:symmetry}, shows that an electric field $E$ along the $x$-direction yields nonzero $\expect{s_{y}}(t)$ and $\expect{l_{y}}(t)$ only if RSOC is present: these components signify thus the SEE and the OEE, respectively. Moreover, the longitudinal probability current $\expect{j_{x}}(t)$ is expectably nonzero, but there is no transversal probability current $\expect{j_{y}}(t)$. However, there are $y$-polarized longitudinal SAM and OAM currents $\expect{j_{x}^{O_y}}(t)$ (i.\,e.\ SAM- and OAM-momentum locking). Besides, we find $x$- and $z$-polarized transversal SAM and OAM currents $\expect{j_{y}^{O_x}}(t)$ and $\expect{j_{y}^{O_z}}(t)$, i.\,e.\ Hall currents carrying SAM and OAM. 

The numerical results presented in what follows fully comply with the above symmetry relations. For the periodic samples discussed here, the mentioned expectation values are homogeneously distributed and vanish without a laser pulse, i.\,e.\ they are induced by laser excitation. In order to quantify the magnitude of the photo-induced signals that arise from the RSOC, we produced numerical results for systems without RSOC for comparison (see Appendix~\ref{sec:noRSOC}). 

\begin{figure*}[ht!]
    \centering
    \includegraphics[width=0.8\textwidth]{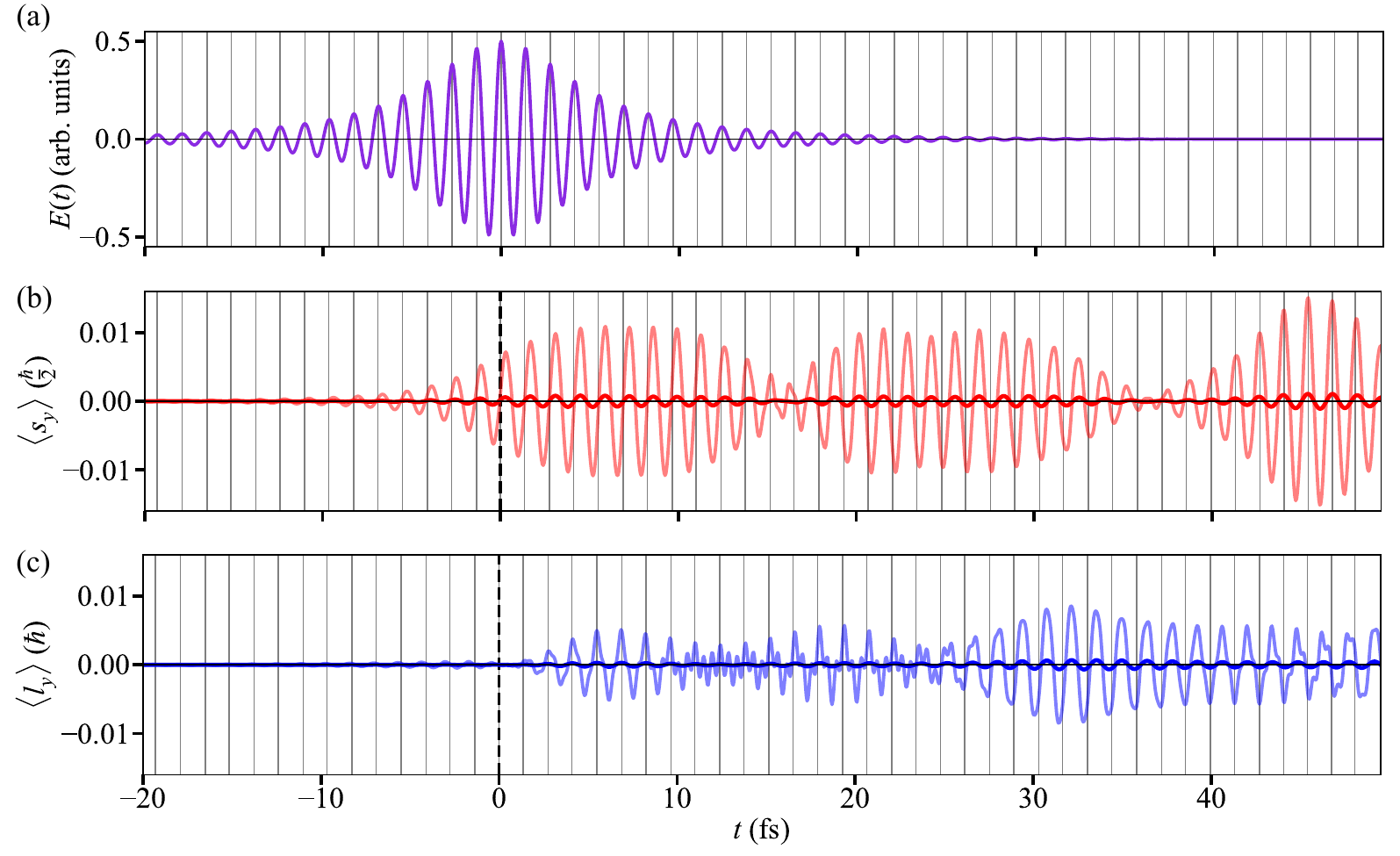}
    \caption{Ultrafast spin and orbital Edelstein effect in a Au(001) monolayer. Panel~(a): electric field of the laser pulse (in arbitrary units). For parameters, see the text. Panels~(b) and (c): photo-induced $y$-components $\expect{s_{y}}(t)$ and $\expect{l_{y}}(t)$ of spin and orbital angular momenta, respectively. Thicker lines in (b) and (c) represent the data convoluted with a Gaussian with standard deviation $\sigma = \unit[10]{fs}$ to visualize the main trends better. Vertical lines indicate the maxima of the laser's electric field.}
    \label{fig:SAM-OAM}
\end{figure*}

\textcolor{black}{In general, the OEE requires broken inversion symmetry~\cite{johansson2024}. However, unlike the SEE, it does not necessarily rely on spin–orbit interaction. In periodic systems, an orbital texture can arise, for instance, in chiral materials~\cite{yoda2018, goebel2025}, where inversion symmetry is inherently broken, and energy splitting of spin-degenerate bands is not required. Such orbital textures---often antisymmetric---can result from orbital hybridization~\cite{go2017, go2018} or be described using the modern framework, which also captures topological aspects~\cite{xiao2006, shi2007}.}

\textcolor{black}{In the case of the Au film studied here, the intrinsic crystal structure is inversion symmetric. However, inversion symmetry is effectively broken by the RSOC term in the Hamiltonian, which can emerge due to the influence of a substrate. This scenario is analogous to the conventional Rashba model, where inversion symmetry breaking gives rise to the SEE. In our system, however, excitation is induced by a laser pulse, and the dipole-transition matrix elements play a crucial role. This necessitates the inclusion of atomic SOC to facilitate the required orbital mixing. Consequently, both SEE and OEE appear in the system.}

\subsection{Laser-induced nonlinear Edelstein effects}\label{subsec:nonlinear_Edelstein_effects}
The photo-induced SAM and OAM components do not follow the laser intensity; their envelope differs from a Lorentzian shape (cf. Figure~\ref{fig:SAM-OAM}). The induced moments do not scale with the laser's electric field amplitude and thus correspond to a nonlinear Edelstein effect~\cite{ye2024}. Supporting this viewpoint, additional simulations confirm that the spin and orbital responses in our system exhibit clear nonlinear dependence on the amplitude of the driving electric field. Accordingly, we interpret the laser-induced signals in Figure~\ref{fig:SAM-OAM} as manifestations of nonlinear SEE and OEE, respectively. 

In the data, pronounced maxima appear at about $\unit[7.5]{fs}$ (panels~b and~c), which is clearly after the laser's maximum at $\unit[0]{fs}$. Moreover, both signals exhibit distinct beating patterns after the laser pulse has faded away. In a recent publication, we related the \textcolor{black}{time-resolved} response to the laser field (orbital Hall currents in that case~\cite{busch2024}) to the classical \textcolor{black}{dynamics} analog of a driven \textcolor{black}{damped} oscillator. \textcolor{black}{Such a model is able to qualitatively capture characteristic phase relations (notably the phase shift of $\pi/2$ between the orbital Hall currents and the driving field in Ref.~\onlinecite{busch2024}).} 
\textcolor{black}{This motivates comparing the beating patterns observed in Fig.~\ref{fig:SAM-OAM} to those in classical systems, such as a driven pendulum at low amplitudes~\cite{Cross_2021}. For the latter, beating is observed between the transient and steady-state response. However, we stress that this analogy serves an illustrative purpose and should not be viewed as a rigorous explanation of the microscopic quantum dynamics.}

\textcolor{black}{
Regarding the possible origins of the observed beating patterns---particularly whether multiphoton processes are involved---we note that the applied laser pulse is dominated by its carrier frequency [cf.~Figure~\ref{fig:FFT}(a)]. In our computational framework, \textsc{evolve}, the laser's electromagnetic field is treated classically, rendering the concept of photons---quantized excitations of the field---strictly speaking inapplicable. 
However, the simulations reveal a nonlinear dependency of the response amplitudes on the driving electric field. Such nonlinear behaviour is sometimes associated with multiphoton physics, although the classical treatment does not capture photon-based processes.}

\textcolor{black}{
Consistent with earlier findings reported in Ref.~\onlinecite{Toepler2021}, we observe that electronic states far above the chemical potential---at energies several times larger than the laser’s carrier energy---become transiently populated. This occurs through the continued evolution of the `excited' density matrix $\rho(t)$, as governed by the von Neumann equation. Both excitation and de-excitation processes persist throughout the duration of the laser pulse. Since the induced population remains nearly unchanged after the pulse (see Figure~1 in Ref.~\onlinecite{Toepler2021}), the beating patterns reflect coherent post-pulse dynamics of the excited electronic states rather than ongoing transitions. 
}

The signals do not vanish after the laser pulse. However, maxima and minima follow rapidly after the pulse (a frequency analysis is given in Appendix~\ref{sec:frequencyAnalysis}), so the time average is very small. This feature presents experiments with the problem of detecting the laser-induced SAM or OAM: it requires temporal resolution on the femtosecond timescale (alternatively, the modulus of the laser-induced signals might be measured). One may consider a time average over several periods in order to detect trends in the signal's evolution. A convolution of the original oscillating data with a Gaussian with a standard deviation of $\unit[10]{fs}$ shows that these time averages are tiny as mentioned before (thicker lines in Figure~\ref{fig:SAM-OAM}); this finding implies that there is almost no net SAM and OAM induced by the laser pulse.
\begin{figure*}[ht!]
    \centering
    \includegraphics[width=0.85\textwidth]{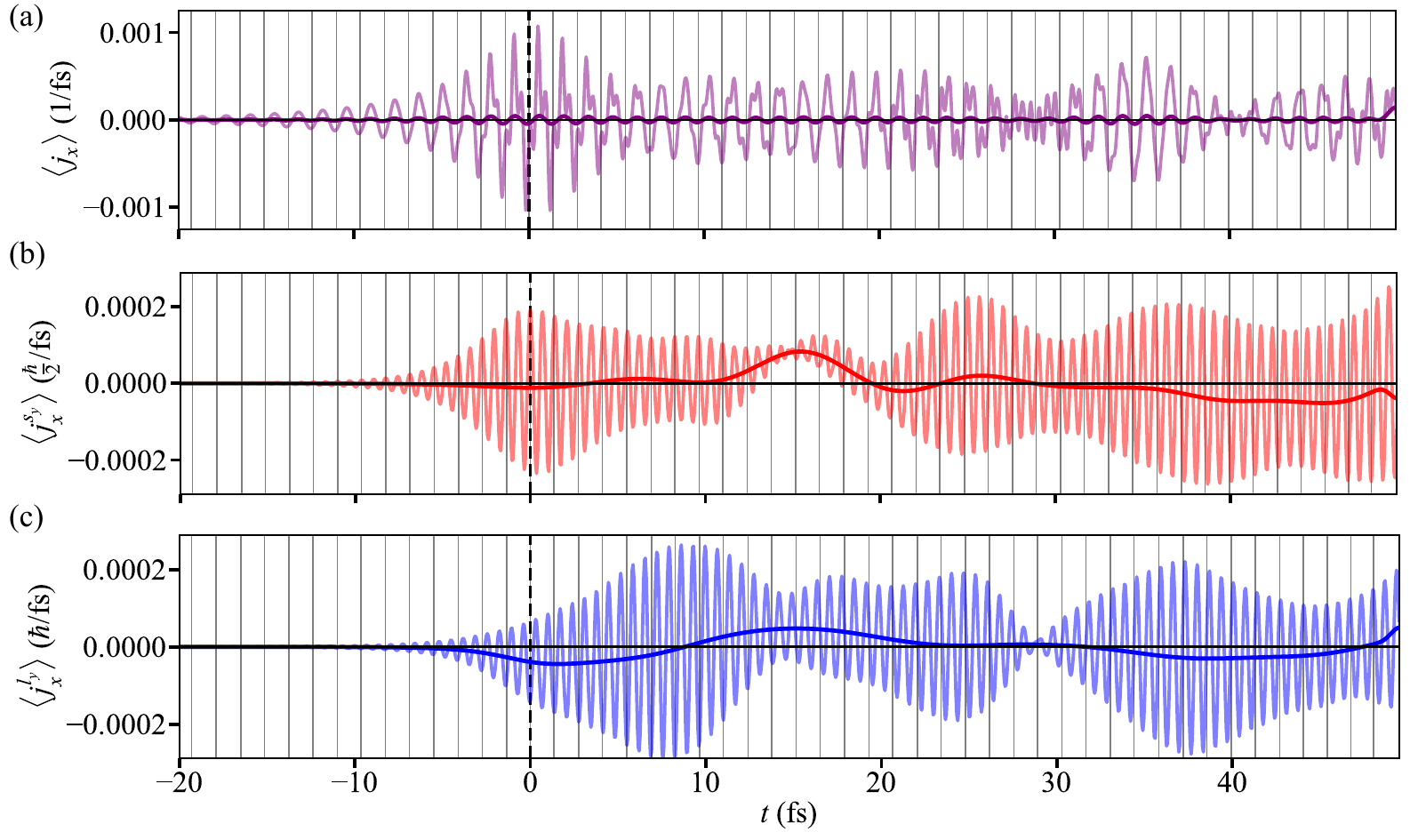}
    \caption{Photo-induced longitudinal currents in a Au(001) monolayer. Panel~(a): probability current $\expect{j_{x}}(t)$ between next-nearest neighbor sites in $x$-direction. Panels~(b) and (c): as panel~(a) but for the $y$-polarized spin and orbital angular momentum currents $\expect{j_{x}^{s_y}}(t)$ and $\expect{j_{x}^{l_y}}(t)$, respectively. Like in Figure~\ref{fig:SAM-OAM}, thicker lines represent the data convoluted with a Gaussian, and vertical lines indicate the maxima of the laser's electric field.}
    \label{fig:longitudinal-currents}
\end{figure*}

In the present simulations, the coupling of the electron system to a bosonic heat bath via Lindblad operators has been neglected \cite{Toepler2021, Ziolkowski23}, as our primary focus is on the fundamental processes that are induced by the laser excitation. Nonetheless, we performed additional calculations that include this coupling. As expected, the key processes discussed here still appear. The magnitude and overall features (e.g., rapid oscillations and beating patterns) of the resulting signals are slightly modified within the time range shown in the figures. Consistent with expectations, the presence of the heat bath leads to a reduction in signal amplitude following the pulse, occurring within a few hundred femtoseconds, depending on the coupling parameters.

Within the \textsc{evolve} framework, only the intrinsic contributions to the Edelstein effects are considered, as the system is assumed to be free from defects and thus lacking any extrinsic contributions. This naturally raises the question: why do Edelstein effects manifest on ultrafast timescales?

Since the laser-induced magnetic moments depend nonlinearly on the laser's electric field, the intrinsic Edelstein effect is nonlinear and distinct from the linear extrinsic Edelstein effect that is typically discussed in a Rashba system. In the context of linear response theory, as described by the Kubo formalism, the emergence of a linear extrinsic Edelstein effect is closely tied to the symmetries of time reversal $\mathcal{T}$ and spatial inversion $\mathcal{I}$. 
A broken $\mathcal{I}$ symmetry is necessary for both extrinsic and intrinsic contributions, ensured here by the out-of-plane potential gradient (Appendix~\ref{sec:symmetry}). However in linear response, a nonvanishing intrinsic contribution also requires the breaking of $\mathcal{T}$ symmetry, even in a nonmagnetic system. Thus, \textcolor{black}{within our model, } the breaking of time-reversal symmetry \textcolor{black}{would be} essential for the \textcolor{black}{linear intrinsic} Edelstein effect, as detailed in Ref.~\onlinecite{johansson2024}.

\begin{figure*}
    \centering
    \includegraphics[width=0.85\textwidth]{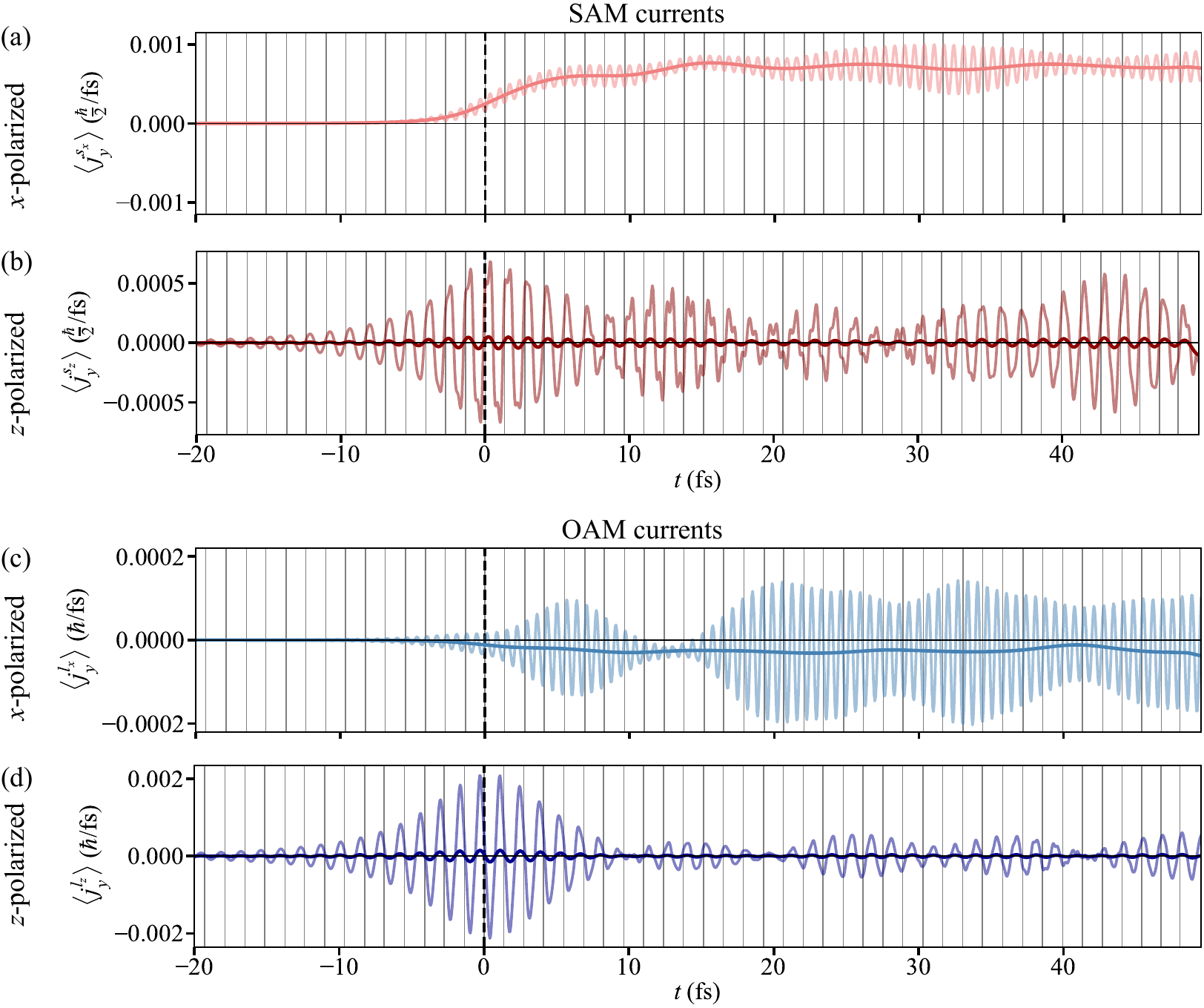}
    \caption{Photo-induced transversal currents of spin and orbital moment in an Au(001) monolayer. Panels (a) and (b): $x$- and $z$-polarized currents of spin angular momentum $\expect{j_{y}^{s_x}}(t)$ and $\expect{j_{y}^{s_z}}(t)$, respectively. Panels (c) and (d), the respective $x$- and $z$-polarized orbital angular momentum currents $\expect{j_{y}^{l_x}}(t)$ and $\expect{j_{y}^{l_z}}(t)$ are shown, as indicated. Like in Figure~\ref{fig:SAM-OAM}, thicker lines represent the data convoluted with a Gaussian, and vertical lines indicate the maxima of the laser's electric field.}
    \label{fig:transversal-currents}
\end{figure*}

A nonlinear Edelstein effect\, as we observe here, can occur independently of time-reversal symmetry. This was explicitly demonstrated by Vignale and Tokatly in the case of a noninteracting, ideal two-dimensional electron gas (2DEG)~\cite{vignale2016}. Laser-induced nonlinear Edelstein effects in nonmagnetic systems have previously been attributed to energy dissipation associated with photocurrents generated during laser excitation~\cite{xu2021}. When the laser pulse interacts with the system, it injects energy and couples electronic states that were previously disconnected. This results in a redistribution of population among the energy levels: the density matrix becomes more complex, acquiring additional nonzero elements both on and off its diagonal. 

\subsection{Laser-induced currents}
\label{subsec:laser_induce_currents}
As mentioned above, the ultrafast Edelstein effects are accompanied by laser-induced currents. The longitudinal probability current (panel a in Figure~\ref{fig:longitudinal-currents}) mainly follows the laser's oscillation `under the pulse', but is slightly offset; its envelope is maximal a bit after the pulse maximum. The presence of other frequencies yields a beating pattern `after the pulse' (see Appendix~\ref{sec:frequencyAnalysis}).

The longitudinal $y$-polarized SAM current (panel b in Figure~\ref{fig:longitudinal-currents}) exhibits a frequency that is twice as large as that of the laser pulse. This feature has been observed for the ultrafast orbital Hall effect~\cite{busch2024}; it is explained in a ``two-current model'': the sign of the laser-induced $y$-polarized SAM is strictly related to the sign of the laser's electric field (confer Appendix~\ref{sec:symmetry}); this also holds for the probability current. In one half-period of the laser oscillation, positive $y$-polarized SAM is transported along the $+x$ direction. In the next half-period, negative $y$-polarized SAM is transported along $-x$; the latter is equivalent to the transport of positive $y$-polarized SAM along $+x$. This results in the doubled frequency of the longitudinal $y$-polarized SAM current. The same argument holds for the OAM current (panel c in Figure~\ref{fig:longitudinal-currents}).
Both SAM and OAM currents exhibit a beating pattern. However, in contrast to $\expect{s_{y}}(t)$ and $\expect{l_{y}}(t)$ (Figure~\ref{fig:SAM-OAM}), their average is not small. 

Having addressed the angular momenta and (longitudinal) currents in the foregoing, we now discuss the transversal currents brought about by the laser excitation (Figure~\ref{fig:transversal-currents}). The $x$-polarized currents $\expect{j_{y}^{O_x}}(t)$ exhibit different time signatures (panels a and c). The most striking feature is that the SAM current increases with time until a plateau is reached (panel a). By contrast, the OAM current is rapidly oscillating about a very small average value (panel c).

Also the $z$-polarized transversal currents $\expect{j_{y}^{O_z}}(t)$ deviate significantly from each other. While the SAM current (panel b) is sizable `after the laser pulse' (as is seen also for $\expect{j_{y}^{s_x}}(t)$ in panel a), the respective OAM current (panel d) is largest in magnitude during the pulse and smaller afterwards. 

Note that only the $z$-polarized transversal currents as well as the longitudinal probability current exist even in the absence of RSOC (cf. Figure~\ref{fig:currents-noSOC} in Appendix~\ref{sec:noRSOC}).

As mentioned above, we performed our simulations without coupling the electron system to a bosonic heat bath. If the latter is included, the laser-induced currents would also decay after the illumination. On timescales of at least several hundred femtoseconds after the laser pulse, the signal amplitudes would be reduced until they vanish eventually, since the system relaxes towards an equilibrium state.

The transversal currents depicted in Figure~\ref{fig:transversal-currents} substantiate that SAM and OAM currents behave differently, which may be attributed to the very nature of SAM and OAM: SAM is an intrinsic feature of electrons, while OAM is motion-related and determined by the lattice geometry as well as by the hybridization of the (occupied) orbitals (cf.\ the ACA~\cite{go2017, go2018, go2021}).

Additionally, a frequency analysis can help to determine the origin of the currents. Specifically, RSOC-induced currents [$\expect{j_{x}^{O_y}}(t)$ and $\expect{j_{y}^{O_x}}(t)$] oscillate at even multiples of the laser frequency, whereas currents [$\expect{j_{x}}(t)$ and $\expect{j_{y}^{O_z}}(t)$] that exist without RSOC (but with atomic spin-orbit coupling) oscillate at odd multiples of $\omega$ (for details, see Appendix~\ref{sec:frequencyAnalysis}).

Finally, we would like to note that it might be counterintuitive, at first glance, that there are nonzero polarized currents but zero polarizations; recall that the $x$- and $z$-components of the angular momenta vanish (see Appendix~\ref{sec:symmetry}), while the associated transversal currents do not (Figure~\ref{fig:transversal-currents}). An explanation relies again on the two-current model: a positive polarized current can be viewed either as transport of positive angular momentum in one direction or as transport of negative angular momentum in the opposite direction (say, $j_{y}^{\uparrow} \equiv j_{-y}^{\downarrow}$). If both currents have the same magnitude (in absolute value), there is a positive net polarized current, but the polarization vanishes everywhere at any time.

\section{Concluding remarks}
\label{sec:conclusion} 
In this study, our simulations with \textsc{evolve} are in full agreement with symmetry considerations and demonstrate that RSOC manifests itself in the existence of nonlinear spin and orbital Edelstein effects induced by a femtosecond laser pulse.
Thereby, we extend the well-known steady-state Edelstein effects to the femtosecond timescale, which advances our understanding of ultrafast light-matter interactions.
In contrast to our findings for the ultrafast orbital Hall effect, which is largest `under the laser pulse', the ultrafast Edelstein effects are still seizable after the laser pulse has decayed. 

Additionally, our simulations show that the RSOC-induced ultrafast Edelstein effects are accompanied by both longitudinal and transversal currents, which offer additional transport channels.
Our findings revealed significant differences in the mechanisms governing the transversal transport of spin and orbital angular momenta. These insights deepen our comprehension of angular momentum dynamics and may pave the way for alternative approaches in spintronics and orbitronics at ultrafast timescales.

This investigation focuses on an Au(001) monolayer, where RSOC may be induced by an insulating substrate. By adopting this deliberately simplified model, we aim to identify the key features of the spin and orbital Edelstein effects. However, the inherent simplifications highlight the need for further studies using more realistic and experimentally relevant samples. Future research should explore a wider range of materials and incorporate additional factors, such as structural imperfections and substrate interactions, to bridge the gap between theoretical models and real-world applications.

\acknowledgments
This work is funded by the Deutsche Forschungsgemeinschaft (DFG, German Research Foundation)---Project-ID 328545488---TRR~227, project~B04. 
This work was supported by the EIC Pathfinder OPEN grant 101129641 ``Orbital Engineering for Innovative Electronics''.

\subsection*{Data Availability Statement}
The data that support the findings of this article are openly available~\cite{data}.

\appendix

\section{Symmetry analysis}
\label{sec:symmetry}
As in the main text, we consider a 2D sample in the $ xy$ plane. For the RSOC, we address its symmetry properties by the potential gradient in $z$-direction, $\partial V / \partial z$, rather than dealing with the associated TB parameters. The electric field $E$ of the laser is along the $x$-axis, and probability currents flow in $x$- ($j_{x}$, longitudinal) or in $y$-direction ($j_{y}$, transversal).

The introduction of two edges along the $y$-direction produces the geometry of a nanoribbon. Taking the Cartesian $x$-axis as the central line along the nanoribbon, the sample is separated into a left ($\half{l}$) and a right ($\half{r}$) half, both of which have to be taken into account when applying the symmetry operations. The symmetry operations and their effect on the quantities are listed in Table~\ref{tab:symmetry}. Note that the results obtained for the spin polarization $\vec{s} = (s_{x}, s_{y}, s_{z})$ also apply to the orbital polarization $\vec{l} = (l_{x}, l_{y}, l_{z})$.   

\paragraph*{2D system without edges.}
Without an electric field (no laser\textcolor{black}{, i.\,e.\,when the system is in thermal equilibrium}) only operations that leave the Rashba term $\partial_{z} V$ invariant are considered: $\operator{1}$, $\operator{C}_{2z}$, $\operator{m}_{yz}$, $\operator{m}_{zx}$. As a result, both spin polarization and the currents vanish. \textcolor{black}{Note that this statement only holds for the ground state scenario before the laser pulse illuminates the sample. After the laser pulse has vanished again, the system is still in a laser-excited nonequilibrium state and not in thermal equilibrium.}

With electric field \textcolor{black}{(with laser, i.\,e.\ when the system is in a nonequilibrium state)}, considering only operations that leave the Rashba term and the electric field invariant ($\operator{1}$, $\operator{m}_{zx}$) yields that $s_{y}$ is nonzero: this is the Rashba-Edelstein effect (without Rashba contribution to the spin-orbit interaction, there is no Rashba-Edelstein effect because of the then allowed additional operations $\operator{C}_{2x}$ and $\operator{m}_{xy}$\textcolor{black}{; we note in passing that the laser-induced $s_y$ persists after the laser pulse has vanished since there is no damping mechanism considered in the simulations).} Moreover, the longitudinal probability current $j_{x}$ is nonzero, but there is no transversal probability current $j_{y}$. By combining the symmetry properties of spin-polarization components and currents we conclude that there is a $y$-spin-polarized longitudinal current $j_{x}^{y}$ as well as $x$- and $z$-spin-polarized transversal currents $j_{y}^{x}$ and $j_{y}^{z}$. As explained above, the symmetry analysis implies that a $y$-polarized longitudinal orbital current as well as $x$- and $z$-polarized transversal orbital currents are allowed.

\begin{table}
	\centering
	\caption{Symmetry operations on a 2D sample and their effect on the potential gradient $\partial_{z} V$, the laser's electric field $E$ along the $x$-direction, the spin-polarization vector $\vec{s} = (s_{x}, s_{y}, s_{z})$, and the probability currents $j_{x}$ and $j_{y}$ in $x$- and $y$-direction, respectively. $\operator{1}$ is the identity operation, rotations about $\pi$ are labeled $\operator{C}_{2\mu}$ ($\mu = x, y, z$ rotation axis), reflections are $\operator{m}_{\mu\nu}$ ($\mu$ and $\nu$ specify the reflection plane). 
    For a nanoribbon, the left and the right half have to be considered ($\half{l}$ and $\half{r}$; see text); these labels can be ignored when analyzing a 2D system without edges.}
\begin{tabular}{c|r|r|rrr|rr}
	\hline \hline
	$\operator{1}$ & $\partial_z V$ & $E$ & $s_{x}^\half{r}$ & $s_{y}^\half{r}$ & $s_{z}^\half{r}$ & $j_{x}^\half{r}$ & $j_{y}^\half{r}$ \\
	$\operator{C}_{2x}$ & $-\partial_z V$ & $E$ & $s_{x}^\half{l}$ & $-s_{y}^\half{l}$ & $- s_{z}^\half{l}$ & $j_{x}^\half{l}$ & $-j_{y}^\half{l}$ \\
	$\operator{C}_{2y}$ & $-\partial_z V$ & $-E$ & $-s_{x}^\half{r}$ & $s_{y}^\half{r}$ & $ -s_{z}^\half{r}$  & $-j_{x}^\half{r}$ & $j_{y}^\half{r}$ \\
	$\operator{C}_{2z}$ & $\partial_z V$ & $-E$ & $-s_{x}^\half{l}$ & $-s_{y}^\half{l}$ & $s_{z}^\half{l}$  & $-j_{x}^\half{l}$ & $-j_{y}^\half{l}$ \\
	$\operator{m}_{xy} $ & $-\partial_z V$ & $E$ & $-s_{x}^\half{r}$ & $-s_{y}^\half{r}$ & $s_{z}^\half{r}$  & $j_{x}^\half{r}$ & $j_{y}^\half{r}$ \\
	$\operator{m}_{yz} $ & $\partial_z V$ & $-E$ & $s_{x}^\half{r}$ & $-s_{y}^\half{r}$ & $-s_{z}^\half{r}$  & $-j_{x}^\half{r}$ & $j_{y}^\half{r}$ \\
 	$\operator{m}_{zx} $ & $\partial_z V$ & $E$ & $-s_{x}^\half{l}$ & $s_{y}^\half{l}$ & $-s_{z}^\half{l}$  & $j_{x}^\half{l}$ & $-j_{y}^\half{l}$ \\ \hline \hline
\end{tabular}
    \label{tab:symmetry}
\end{table}

\paragraph*{Nanoribbon.} For completeness and in view of experiments in which samples are unavoidably finite, we briefly analyze a nanoribbon.

Without an electric field, $s_{x}$ may be nonzero but with opposite signs in opposite halves ($s_{x}^{\half{l}} = -s_{x}^{\half{r}}$). The longitudinal probability current vanishes in the entire ribbon. However, $j_{y}^{\half{l}} = -j_{y}^{\half{r}}$ is allowed for the transversal current. Since there is no electric field that drives this transversal current, it is regarded as a persistent current and has to be subtracted from the laser-induced time-dependent current.

For the nanoribbon with electric field, we obtain $s_{\mu}^{\half{l}} = -s_{\mu}^{\half{r}}$ for $\mu = x, z$ and the Rashba-Edelstein effect deduced for the 2D sample without edges. The longitudinal \textcolor{black}{probability and the $y$-polarized currents are} nonzero in the entire ribbon \textcolor{black}{and obey the relations $j_{x}^{\half{l}} = +j_{x}^{\half{r}}$ and $j_{x}^{y \half{l}} = +j_{x}^{y\half{r}}$, but there are longitudinal $x$- and $z$-polarized currents with opposite sign of the spin polarization in opposite halves ($j_{x}^{\mu \half{l}} = -j_{x}^{\mu \half{r}}$, $\mu = x, z$). Similarly, for the transversal probability and $y$-polarized current $j_{y}^{\half{l}} = -j_{y}^{\half{r}}$ and $j_{y}^{y \half{l}} = -j_{y}^{y \half{r}}$ holds. However, for the other two transversal spin-polarized currents it is $j_{y}^{\mu \half{l}} = +j_{y}^{\mu \half{r}}$ for $\mu = x, z$}.

\begin{figure*}[ht!]
    \centering
    \includegraphics[width=0.8\textwidth]{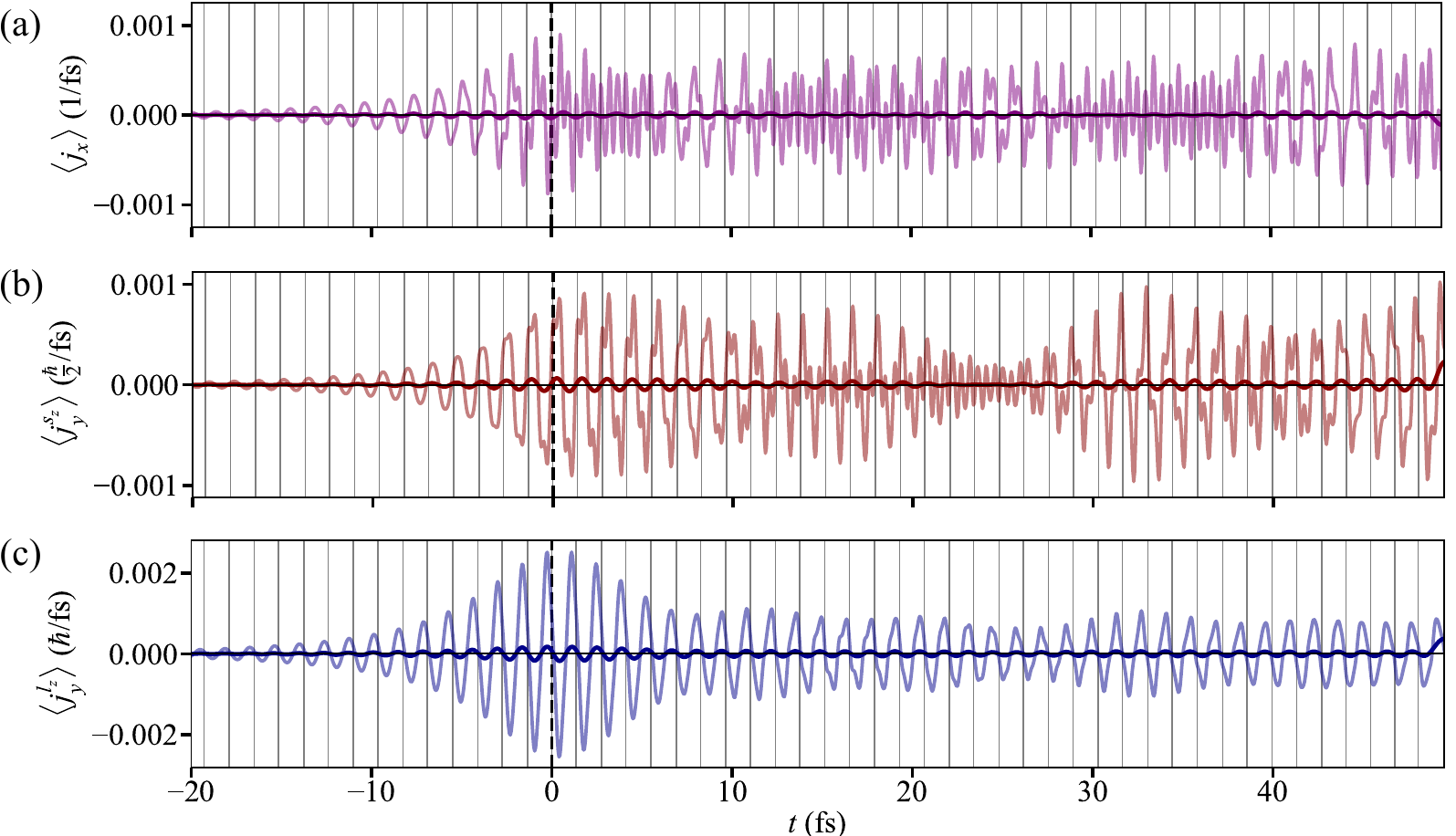}
    \caption{Laser-induced currents in a Au(001) sample without Rashba spin-orbit coupling. Panel~(a): longitudinal probability current $\expect{j_{x}}(t)$ between next-nearest neighbor sites in $x$-direction. Panels~(b) and (c): transversal $z$-polarized spin and orbital angular momentum currents in $y$-direction $\expect{j_{y}^{s_z}}(t)$ and $\expect{j_{y}^{l_z}}(t)$, respectively. Like in Figure~\ref{fig:SAM-OAM}, thicker lines represent the data convoluted with a Gaussian, and vertical lines indicate maxima of the laser's electric field.}
    \label{fig:currents-noSOC}
\end{figure*}

\section{Rashba-type spin-orbit coupling in real-space tight-binding form}
\label{sec:Rashba-SOC-TB}
The Hamiltonian for spin-orbit coupling 
\begin{align}
	\operator{H}_{\mathrm{so}} & = \frac{1}{4 c^{2}} \vec{\sigma} \cdot (\nabla V \times \operator{\vec{p}})
    \label{eq:SOC}
\end{align}
(in Hartree atomic units; $c$ speed of light, $\vec{\sigma}$ vector of Pauli matrices, $V(\vec{r})$ potential, $\operator{\vec{p}} = -\cone \nabla$ momentum operator) describes, when considering a homogeneous potential gradient in $z$-direction, RSOC~\cite{bychkov1984}
\begin{align}
	\operator{H}_{\mathrm{so}}^{\mathrm{R}}
	& = \frac{1}{4 c^{2}} \frac{\partial V}{\partial z}  (\sigma_{y} \operator{p}_{x} - \sigma_{x} \operator{p}_{y}).
    \label{eq:SOC-Rashba}
\end{align}
Assuming further that the gradient is constant, the TB matrix elements of this Hamiltonian are essentially given by the matrix elements of the momentum operator.

We express an electron state
\begin{align}
	\Psi(\vec{r}) & = \sum_{\alpha l} c_{\alpha l} \, \phi_{\alpha}(\vec{r} - \vec{R}_{l})
\end{align}
as a sum over orthonormal orbitals $\phi_{\alpha}(\vec{r})$ of type $\alpha$ centered at lattice sites $\vec{R}_{l}$. The action of the momentum operator $\operator{\vec{p}}$ on that state
is
\begin{align}
	\operator{\vec{p}} \Psi(\vec{r})
	& = - \cone \sum_{\alpha l} c_{\alpha l} \nabla \phi_{\alpha}(\vec{r} - \vec{R}_{l}).
\end{align}
Since in semi-empirical TB approaches the explicit form of the orbitals is usually unknown, the gradient cannot be calculated and, therefore, has to be determined otherwise.
The directional derivative of an orbital
\begin{align}
	\nabla \phi_{\alpha}(\vec{r} - \vec{R}_{l})
	& = \frac{\vec{d}}{d} \lim_{d \to 0} \frac{\phi_{\alpha}(\vec{r} + \vec{d} - \vec{R}_{l}) - \phi_{\alpha}(\vec{r} - \vec{R}_{l})}{d},
\end{align}
with respect to a certain direction $\vec{d}$ becomes without taking the limit
\begin{align}
	\nabla \phi_{\alpha}(\vec{r} - \vec{R}_{l})
	& \approx \frac{\vec{d}}{d^{2}} \left[ \phi_{\alpha}(\vec{r} + \vec{d} - \vec{R}_{l}) - \phi_{\alpha}(\vec{r} - \vec{R}_{l}) \right].
\end{align}
`Coarsening' the gradient by taking $\vec{d}$ as a lattice vector~$\vec{R}$ and exploiting the orthonormality of the orbitals, we arrive at the matrix element
\begin{align}
    \braket{\Psi| \operator{\vec{p}}| \Psi} & \approx
	-\cone \sum_{\alpha; k, l} c^{\star}_{\alpha k} c_{\alpha l} \frac{\vec{d}}{d^{2}}
	\left( \delta_{\vec{R}_{k}, \vec{R}_{l} - \vec{d}} - \delta_{\vec{R}_{k}, \vec{R}_{l}} \right)
\end{align}
of the momentum operator. Since the site-diagonal terms are already treated within the atomic $\vec{L} \cdot \vec{S}$ contribution to spin-orbit coupling in the Hamiltonian $\operator{H}_{0}$ (cf. Ref.~\onlinecite{Konschuh2010}), we are left with the site-nondiagonal contributions. The above equation thus reduces to
\begin{align}
    \braket{\Psi| \operator{\vec{p}}| \Psi} & \approx
	-\cone \sum_{\alpha; k \not= l} c^{\star}_{\alpha k} c_{\alpha l} \frac{\vec{R}_{l} - \vec{R}_{k}}{|\vec{R}_{l} - \vec{R}_{k}|^{2}}
 \end{align}
with $\vec{d} = \vec{R}_{l} - \vec{R}_{k}$. 
With this approximation, the matrix elements
\begin{align}
    \braket{\Psi| \operator{H}_{\mathrm{so}}^{\mathrm{R}} | \Psi} & \approx
	\sum_{\alpha, \beta; k \not= l} c^{\star}_{\beta k} t_{\beta k, \alpha l}^{\mathrm{R}} c_{\alpha l}
 \end{align}
of the RSOC introduce the spin- and direction-dependent hopping terms
\begin{align}
	t_{\beta k, \alpha l}^{\mathrm{R}}
	& = \cone\, \vec{\sigma} 
	\cdot \vec{\lambda}_{\alpha}^{\mathrm{R}} \times \frac{\vec{R}_{k} - \vec{R}_{l}}{|\vec{R}_{k} - \vec{R}_{l}|^{2}} \, \delta_{\beta \alpha}, \quad k \not= l,
\end{align}
in the TB Hamiltonian $\operator{H}_{0}$, in addition to the conventional Slater-Koster-type hopping terms. The vector $\vec{\lambda}_{\alpha}^{\mathrm{R}}$ of Rashba parameters comprises in particular the potential gradient, besides some constants. $t_{\beta k, \alpha l}^{\mathrm{R}}$ is antisymmetric and hermitian, that is $t_{\beta k, \alpha l}^{\mathrm{R}} = - t_{\alpha l, \beta k}^{\mathrm{R}} = \left( t_{\alpha l, \beta k}^{\mathrm{R}} \right)^{\star}$.

A Rashba parameter is usually obtained from the parabolic dispersion relation $E(\vec{k})$ of a pair of Rashba-split states~\cite{Ast2007} and hence is valid only for that state pair. In the present paper, the set of Rashba parameters comprised in $\vec{\lambda}_{\alpha}^{\mathrm{R}}$ depends on the orbital-type $\alpha = \mathrm{s}, \mathrm{p}_{x},  \mathrm{p}_{y}$ etc. It could also depend on the pair of sites $k$ and $l$.

For the 2D systems lying within the $xy$-plane that are investigated in the present paper, a constant potential gradient in $z$-direction is considered, which implies $\vec{\lambda}_{\alpha}^{\mathrm{R}} = (0, 0, \lambda_{z \alpha}^{\mathrm{R}})$ and reproduces the well-known combinations of Pauli matrices and directions in
\begin{align}
	t_{\beta k, \alpha l}^{\mathrm{R}}
	& = \cone\, \lambda_{z \alpha}^{\mathrm{R}} \frac{ \sigma_{y} (R_{k x} - R_{l x}) - \sigma_{x} (R_{k y} - R_{l y})}{|\vec{R}_{k} - \vec{R}_{l}|^{2}}  \, \delta_{\beta \alpha}
    \label{eq:RSOCz}
\end{align}
($k \not= l$), as found in Eq.~\eqref{eq:SOC-Rashba} (effects of in-plane potential gradients present at surfaces with three-fold rotational symmetry has been studied in Ref.~\onlinecite{Premper2007}). Moreover, ignoring the orbital dependence, the vector  
\begin{align}
  \vec{n}_{k l} & \equiv \vec{\lambda}^{\mathrm{R}} \times \frac{\vec{R}_{k} - \vec{R}_{l}}{|\vec{R}_{k} - \vec{R}_{l}|^{2}}
\end{align}
lies within the $xy$-plane and is perpendicular to the distance vector between the two sites. It allows to rewrite equation~\eqref{eq:RSOCz} as
$t_{k l}^{\mathrm{R}} = \cone\, \vec{\sigma} \cdot \vec{n}_{k l}$, which is utilized in model Hamiltonians~\cite{chen2014, zhang2018, Zhou2019, busch2020, busch2021}.

\textcolor{black}{In the context of the present study, the relevant angular momentum processes---mechanisms by which angular momentum is generated, transferred, or converted between different degrees of freedom---are the dynamics of SAM and SOC. Concerning SAM dynamics, the SEE arises from the generation of a nonequilibrium spin polarization in response to an applied electric field in systems with SOC~\cite{go2020, johansson2024}\@. This involves the redistribution of SAM through interactions with external fields and scattering processes. Our model explicitly includes this mechanism by treating spin degrees of freedom coupled via SOC\@.}

\textcolor{black}{The second mechanism, SOC, provides a coupling between SAM and OAM, allowing for the interconversion of linear momentum (associated with charge currents) and SAM\@. This interconversion is the central angular momentum transfer mechanism underlying the Edelstein effect~\cite{johansson2021}. In our framework, where the electromagnetic field is treated classically and the lattice remains rigid, SOC plays the dominant role (our model includes both atomic SOC and RSOC). In more comprehensive treatments, additional channels such as OAM of electrons or angular momentum transfer to the lattice---via phonons or rigid-body rotations---may contribute. However, such processes are often neglected in effective models focused on capturing the essential physics of spintronic effects. Our model does not incorporate these mechanisms, as they are not required for describing the SEE in its fundamental form.}

\section{Laser-induced currents without Rashba spin-orbit coupling}
\label{sec:noRSOC}
In order to reveal the effect of RSOC on the dynamics we performed calculations as discussed in the main text (Section~\ref{sec:results}) with keeping the atomic spin-orbit interaction but without RSOC\@. In accordance with the symmetry analysis in Section~\ref{sec:symmetry}, all laser-induced components of SAM and OAM vanish in the sample without RSOC\@. This means that nonzero SAM and OAM clearly indicate (ultrafast) Rashba-Edelstein effects.

Concerning the laser-induced longitudinal currents, there is only a longitudinal probability current, but neither a SAM nor an OAM current. Moreover, both the transversal $x$-polarized SAM and OAM currents vanish without RSOC, and only the $z$-polarized ones remain. This is in agreement to our previous work on the ultrafast orbital Hall effect in metallic nanoribbons~\cite{busch2024}.

The overall shape of the current signals agrees fairly well with those computed for the sample with RSOC (compare Figures~\ref{fig:longitudinal-currents} and~\ref{fig:transversal-currents} with Figure~\ref{fig:currents-noSOC}). Also, the overall magnitude of the signals is barely affected by neglecting RSOC\@.

If not only RSOC but also the atomic spin-orbit coupling is neglected, the $z$-polarized transversal SAM current also vanishes, whereas the $z$-polarized transversal OAM current within the ACA is still nonzero. These findings are in agreement with literature; Go \textit{et al.} demonstrated that within the ACA, hybridization of specific orbitals gives rise to the orbital Hall effect, which is allowed in nonmagnetic centrosymmetric systems even in the absence of (atomic) spin-orbit interaction~\cite{go2018, go2021}. Yet, the latter is required for the spin Hall effect in such systems.  

\begin{figure}[ht!]
    \centering
    \includegraphics[width = 0.95\columnwidth]{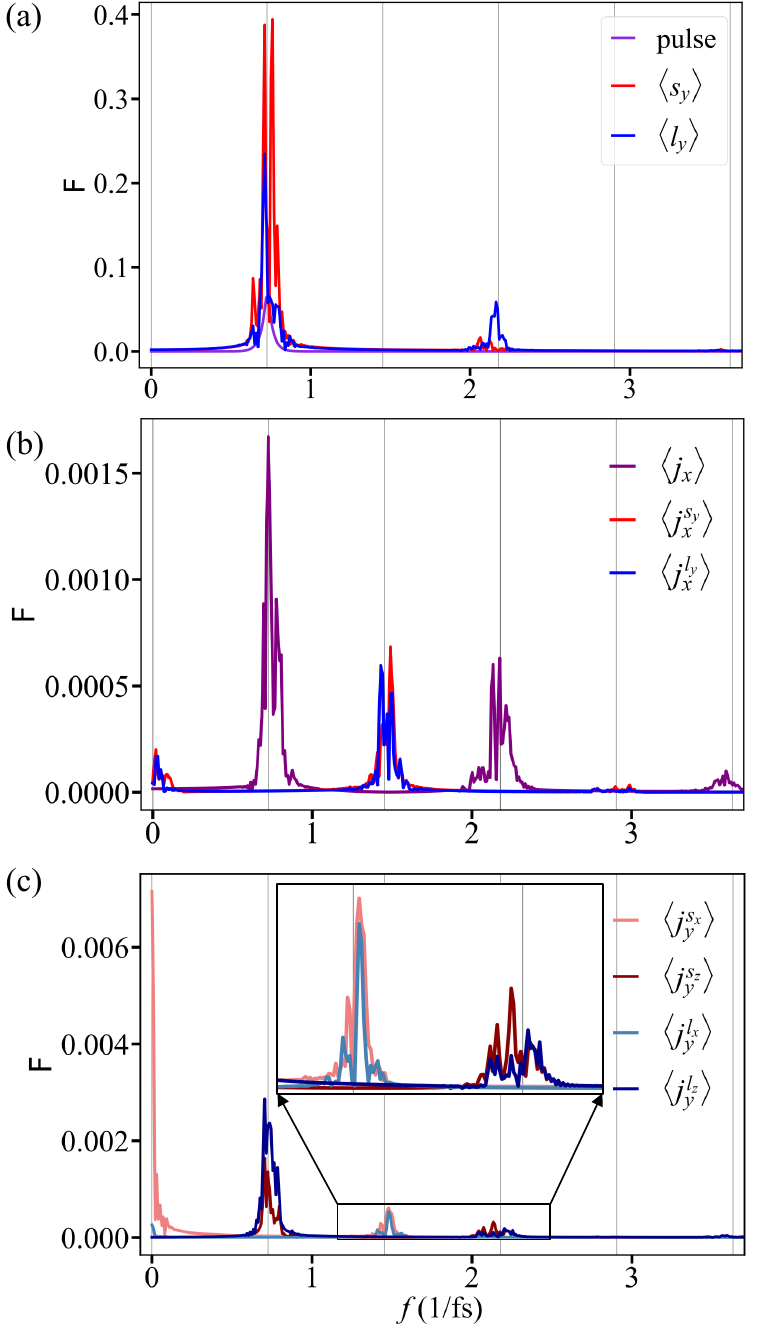}
    \caption{Frequency analysis. (a) Fast Fourier transformations of the laser pulse and the induced $y$-components of the spin and the orbital angular momentum\@. (b) As (a), but for the longitudinal probability, spin, and orbital angular momentum currents flowing in $x$-direction. (c) As (b), but for the respective transversal currents flowing in $y$-direction. \textcolor{black}{The inset in panel (c) shows the vicinity of $2\,\omega$ and $3\,\omega$ for better visibility.} Gray vertical lines indicate multiples of the laser frequency.}
    \label{fig:FFT}
\end{figure} 

\section{Frequency analysis}
\label{sec:frequencyAnalysis}

Fast Fourier transformations (FFT) of the time-dependent signals (presented in Figures~\ref{fig:SAM-OAM}, \ref{fig:longitudinal-currents}, and \ref{fig:transversal-currents}) reveal noticeable differences regarding their origin.

The SAM and OAM components display strong FFT maxima at about the laser's frequency and at about higher odd multiples of $\omega$ [Figure~\ref{fig:FFT}(a); $\omega$, $3\, \omega$, $5\, \omega$, \ldots], as does the longitudinal current $\expect{j_{x}}$. This finding supports that these SAM and OAM components are brought about by the perturbation (i.\,e.\ they are nonequilibrium features).

The respective longitudinal $y$-polarized SAM and OAM currents, $\expect{j_{x}^{s_{y}}}$ and  $\expect{j_{x}^{l_{y}}}$, have maxima concentrated at even multiples of the laser frequency [Figure~\ref{fig:FFT}(b); $0$, $2\, \omega$, $4\, \omega$, \ldots]. This frequency doubling with respect to the laser-induced SAM and OAM components is explained in a two-current model as follows. The sign of the laser-induced angular momentum is directly related to the orientation of the laser's electric field, giving an oscillation with $\omega$. The longitudinal current also oscillates with $\omega$ and transports angular momentum in $+x$ direction in one half-period of the laser pulse and the opposite angular momentum along $-x$ in the next half-period. Since the transport of angular momentum in one direction is equivalent to the transport of the opposite angular momentum in the opposite direction, the SAM and OAM currents oscillate with $2\, \omega$. Such an effect has also been seen in the laser-induced ultrafast orbital Hall effect~\cite{busch2024}. The zero-frequency features indicate sustained, small contributions.

Concerning the transversal SAM and OAM currents \textcolor{black}{[Figure~\ref{fig:FFT}(c)]}, we find that the $x$-polarized ones, $\expect{j_{y}^{s_{x}}}$ and $\expect{j_{y}^{l_{x}}}$, show maxima at even multiples of $\omega$. Recall that these are brought about by the RSOC, as are the longitudinal currents $\expect{j_{x}^{s_{y}}}$ and $\expect{j_{x}^{l_{y}}}$. 

By contrast, the $z$-polarized transversal SAM and OAM currents, $\expect{j_{y}^{s_{z}}}$ and $\expect{j_{y}^{l_{z}}}$, have FFT maxima at odd multiples of $\omega$. These currents show up even if RSOC is switched off in the simulations (Figure~\ref{fig:currents-noSOC}). This finding suggests that a frequency analysis allows one to conclude on the origin of the currents. In the present case, currents brought about by RSOC oscillate with even multiples of the laser's frequency, while those not associated with RSOC oscillate with odd multiples of $\omega$.

\bibliography{references}

\end{document}